\documentclass[usenatbib]{mnras}
\usepackage{ae, amsmath, amssymb}

\title[Unexpected LIGO events and M-World]{Unexpected LIGO events and the Mirror World}

\author[Revaz Beradze and Merab Gogberashvili]
{Revaz Beradze,$^{1}$\thanks{E-mail: revazberadze@gmail.com}
and Merab Gogberashvili$^{1,2}$\thanks{E-mail: gogber@gmail.com}
\\
$^{1}$Javakhishvili Tbilisi State University, 3 Chavchavadze Avenue, Tbilisi 0179, Georgia \\
$^{2}$ Andronikashvili Institute of Physics, 6 Tamarashvili Street, Tbilisi 0177, Georgia}

\begin{document}

\maketitle

\begin{abstract}
We consider the possibility that LIGO events GW190521, GW190425 and GW190814 may have emerged from the mirror world binaries. Theories of star evolution predict so called upper and lower mass gaps and masses of these merger components lie in that gaps. In order to explain these challenging events very specific assumptions are required and we argue that such scenarios are order of magnitude more probable in mirror world, where star formation begins earlier and matter density can exceed 5 times the ordinary matter density.
\end{abstract}

\begin{keywords}
Multi-messenger Astronomy -- Black Holes -- Neutron Stars -- Mirror World
\end{keywords}


\section{Introduction}

Since the first direct detection of Gravitational Waves (GW) in 2015 \citep{Abbott:2016blz}, a new era in multi-messenger astronomy began. The first two observing runs of Advanced LIGO/VIRGO detectors observed 10 events of binary Black Hole (BH) merger and one event, GW170817, of binary Neutron Star (NS) coalescence \cite{LIGOScientific:2018mvr}, which still remains the only GW detection accompanied by a gamma-ray counterpart \citep{Goldstein:2017mmi, Savchenko:2017ffs}. Analysis of the data from the first part of the third observing run (O3a) already revealed several interesting and surprising events \citep{Abbott:2020niy}. However, any sign of corresponding electromagnetic radiation, notably also for the events involving NSs, is still absent in the data of O3a \citep{Abbott:2020yvp}.


\subsection{The source from the upper mass gap}

In May 2019, Advanced LIGO/VIRGO detected the signal GW190521, which was radiated by the coalescence of two BHs and is remarkable for a number of reasons:
\begin{itemize}
\item The most distant BH-BH merger (at about 15 billion light-years away);
\item The most massive progenitor BHs ($91.4^{+29.3}_{-17.5} M_\odot$ and $66.8^{+20.7}_{-20.7} M_\odot$);
\item The most massive final BH (at $157.9^{+37.4}_{-20.9} M_\odot$), the very first “intermediate mass BH” ever detected;
\item The greatest amount of mass turned into energy in a single event ($\sim 8 M_\odot$);
\item The shortest-duration definitive signal ever seen (at $\sim 12.7$ milliseconds).
\end{itemize}
These data, initially published in \citep{Abbott:2020tfl, Abbott:2020mjq}, where updated in the second catalog \citep{Abbott:2020niy} and here we present the updated values.

The biggest surprise of GW190521 was that the primary BHs with masses $91 M_\odot$ and $67 M_\odot$ lie within the mass gap produced by pair-instability supernova processes. Current models for evolution of heavy stars predict that temperature in a heavy helium core reaches the point, when production of electron-positron pairs is allowed. So, part of energy of photons that was providing pressure against gravity, is consumed by pair production and star becomes unstable. Stars with helium core mass $32-64 M_\odot$ are subject to pulsational pair instability, decreasing mass by ejection of some amount of matter, and leave remnant with mass less than $65 M_\odot$. Stars with helium mass $65-135 M_\odot$ are affected by pair instability, disrupting entire star and leaving no remnant compact object. Stars with helium core $\gtrsim 135 M_\odot$ are considered to directly collapse to intermediate mass BHs \citep{Heger:2001cd, Woosley:2007qp, Woosley:2016hmi}. So, collapse of a heavy star was unable to produce the primary BHs of GW190521.

Intermediate mass BHs currently are thought to be formed by a hierarchical merger of smaller BHs \citep{Miller:2001ez, Fishbach:2017dwv, Gerosa:2017kvu, Antonini:2018auk, Kimball:2020opk, Mapelli:2020xeq, Liu:2020gif}. In order to coalescence again, initial first generation BHs should be formed in triple or higher multipole systems, or such systems must be assembled in the dense stellar clusters. However, merger product receives a recoil kick from the anisotropic GW emission\citep{Campanelli:2007ew, Lousto:2011kp, Gonzalez:2006md, Brugmann:2007zj, Varma:2018aht}, and it may eject them from clusters and leave unavailable to form new generations of binary BHs \citep{Moody:2008ht, Varma:2020nbm}. Also, even if effective spin parameter of binary BH system was almost zero, BH formed after their merger is characterized by high spin; so spin value may be a good indication for genealogy of BHs \citep{Miller:2002pg, Sedda:2020vwo, Baibhav:2020xdf}. In \citep{Kimball:2020opk}, the analyses of the ten BHs coalescence events from the first two observing runs of LIGO/VIRGO detectors was made and no definite evidence of hierarchical mergers was found.

Merger rate of GW190521-like events was estimated to be $0.13^{+0.30}_{-0.11} ~\rm{Gpc^{-3} yr^{-1}}$ \citep{Abbott:2020mjq}. However, simulations done for nuclear stellar clusters (where the hierarchical mergers are orders of magnitude more common than in globular and young star clusters), yield $10^{-2} - 0.2 ~\rm{Gpc^{-3} yr^{-1}}$ and models with low spin and broad distribution of cluster escape velocities are favored \citep{Mapelli:2020xeq}. In order to obtain the merger rate comparable with GW190521, some optimistic assumptions are required. Particularly, merger rate of GW190521-like events was estimated as \citep{Liu:2020gif},
\begin{equation} \label{rate}
f = f_{\rm{1G}} \times f_{\rm {triple}} \times f_{\rm {survival}} \times f_{\rm {merger}} ~,
\end{equation}
where $f_{\rm{1G}}$ is merger rate of first generation BHs, for which the value from the first two observing runs was used \citep{LIGOScientific:2018mvr},
\begin{equation}
f_{\rm{1G}} \sim (10-100)~ \rm{Gpc^{-3} yr^{-1}}~.
\end{equation}
Assuming a high survival,
\begin{equation}
f_{\rm {survival}} \simeq 60\%~,
\end{equation}
and the large merger fraction,
\begin{equation}
f_{\rm {merger}} \simeq 20\%~,
\end{equation}
together with the admission that formation fraction of stellar triple systems is
\begin{equation}
f_{\rm {triple}} \simeq 50\%~,
\end{equation}
the coalescence rate of GW190521-like events was calculated to be $0.6 - 6 ~\rm{Gpc^{-3} yr^{-1}}$, being in agreement with LIGO/VIRGO estimations, but in the price of some extreme postulations.

Possibility of primordial origin of BHs for the event GW190521 was also considered \citep{Abbott:2020mjq}. But, there are tight constraints on mass distribution of primordial BHs \citep{Sasaki:2018dmp} and some theories of primordial BH formation predict small component spins \citep{Chiba:2017rvs}, that is inconsistent with GW190521. Opportunity of strong gravitational lensing by galaxies or galaxy clusters for the signal GW190521 was also discussed. However, low expecting lensing rate and optical depth, and the absence of multi-image counterpart, disfavor strong lensing hypothesis \citep{Abbott:2020mjq}.


\subsection{The sources from the lower mass gap}

Other unexpected GW signals were the events:
\begin{itemize}
\item GW190425, from the coalescence of objects with the masses $2.0^{+0.6}_{-0.3}M_\odot$ and $1.4^{+0.3}_{-0.3}M_\odot$ \citep{Abbott:2020uma};
\item GW190814, from the merger of a $23.2^{+1.1}_{-1.0} M_\odot$ BH with a $2.59^{+0.08}_{-0.09} M_\odot$ compact object \citep{Abbott:2020khf}.
\end{itemize}
What matters is that the distribution of masses of X-ray binaries reveal apparent so-called lower mass gap $2.5 - 5 M_{\odot}$ between NSs and BHs \citep{Bailyn:1997xt, Ozel:2010su, Ozel:2012ax, Farr:2010tu}. The aforementioned components of LIGO events lie on the edge of that mass gap. Some theoretical models of supernova explosions predict existence of the observed mass gap \citep{Kochanek:2013yca, Pejcha:2014wda}. Nonetheless, some models suggest a smooth transition from NS to BH masses \citep{Woosley:1995ip, Woosley:2002zz, Ertl:2019zks}.

In principle, both components of GW190425 are consistent with being NSs. While the mass of the one component, $1.4^{+0.3}_{-0.3}M_\odot$, falls in a typical range of observed pulsars, the heavier component with the mass $2.0^{+0.6}_{-0.3} M_\odot$ also can be a NS, as far as existence of pulsars with mass of $\sim 2M_\odot$ were confirmed by observations \citep{Demorest:2010bx, Antoniadis:2013pzd}. But there was no optical counterpart to GW190425, unlike to the famous NS-NS merging event GW170817. Withal, with the source-frame chirp mass $1.44 M_\odot$ and the total mass $3.4 M_\odot$, the GW190425 system is significantly massive than any binary NS system observed through electromagnetic radiation.

The most common mechanism for formation of binary NS system is an isolated binary evolution channel (for a review see \citep{Kalogera:2006uj}). Following this path, GW190425 may suggest a population of binary NSs formed in ultra-tight orbits with sub-hour orbital period \citep{Abbott:2020uma}. In order to achieve the system like this, it is required to have phase of mass transfer from post-helium main sequence star onto NS. If the mass ratio between He-star and NS is high enough, common-envelope phase is formed, that would shrink the binary orbit to sub-hour periods \citep{Tauris:2017omb}. If binary survives this common envelope phase, the subsequent supernova kick may be suppressed, as secondary would likely be ultra-stripped \citep{Tauris:2015xra}. The small supernova kick, together with very tight orbital separation, increases the chance for binary to remain bound after supernova. Hierarchical mergers of NSs or mass accretion in active galactic nuclei disks, may also be responsible for the creation of compact objects in the lower mass gap \citep{Yang:2020xyi}. Other possible scenarios, like dynamical formation channel in globular cluster or gravitational lensing of the source of GW190425, is also discussed \citep{Abbott:2020uma}. However, lack of objects discovered with given properties, does not allow to have a definite theory for their formation mechanism for now.

In the case of GW190814, the first component is definitely a BH. But the second object certainly lies in the lower mass gap. It is heavier than the most massive known pulsar and lighter than any BH discovered so far. The mass of the secondary component $2.59^{+0.08}_{-0.09}$  exceeds possible maximum mass allowed for a stable NS in most of the models \citep{Abbott:2018exr, Lim:2019som, Essick:2019ldf}. However, due to theoretical uncertainties the NS-BH scenario cannot be ruled out, e.g. the second component can be a NS in view of some stiffer equation of state of dense nuclear matter \citep{Mueller:1996pm}. It is also possible that it is a quark star in which the original NS was transformed due the fallback of the material after the gravitational collapse of a progenitor star \citep{Berezhiani:2002ks}. Thus, GW190814 can be viewed as BH-NS (or QS) merger which in principle could have an associated GRB and optical counterpart. The probability for the secondary component of GW190814 being exotic compact object, like boson star \citep{Kaup:1968zz}, gravastar \citep{Mazur:2004fk}, strange-quark star \citep{Bombaci:2020vgw} or up-down quark star \citep{Cao:2020zxi} was also considered.

For the completeness note that the secondary of the event GW190814 is presumably a BH, which may have formed by coalescence of NSs, e.g. the remnant of the event GW170817 has the similar mass \citep{LIGOScientific:2018mvr}. But to merge again, hierarchical triple system in the field \citep{Silsbee:2016djf, Antonini:2017ash, Fragione:2019zhm} or in the galactic center \citep{Antonini:2012ad, Petrovich:2017otm, Hoang:2017fvh, Fragione:2018yrb} must be considered. However, with such unequal masses, the abundance of GW190814-like systems is poorly constrained and is not expected to be high.


\section{Sources from mirror world}

With so much confusions arisen during the third observing run, we want to propose a new explanation relating the masses of component compact objects. In \citep{Addazi:2017gne, Beradze:2019dzc, Beradze:2019ujd, Beradze:2019yyp} it was suggested that some GWs detected by LIGO may have emerged from the Mirror World (M-World). M-World is a candidate of Dark Matter (DM) and mirror particles interact with our ordinary world only through gravity. That is why we very rarely see electromagnetic counterpart of GWs (the only one event so far), and as DM density exceeds baryonic matter density 5 times, estimated merger rates are higher than expected in standard scenarios and they agree better with LIGO measurements.

In this article we want to explain the unexpected masses of mergers for recently discovered LIGO events GW190521, GW190425 and GW190814, using M-World scenario. Before that, we shortly review the M-World theory.


\subsection{Mirror world scenario}

Mirror World (M-World) was introduced to restore left-right symmetry of nature, suggesting that each Standard Model particle has its mirror partner with opposite chirality. The fundamental reason for existence of mirror partners has been first revealed in \citep{LeeYang}. Later, based on this idea, theory of M-World was introduced, stating that mirror particles are invisible for ordinary observers and vice versa. Only way for the interaction between these two worlds is gravity. So, GW radiated by mirror matter can be sensed by an ordinary observer. For a review of theoretical foundations of M-World and its cosmological and astrophysical properties see \citep{mirror_okun, mirror_khlopov, mirror}.

If M-World really exists, it was created by the Big Bang along the ordinary universe. But its temperature, $T'$, must be lower than the temperature of our world, $T$. This requirement emerges from the fact that mirror particles, having similar cosmological abundance, also contribute into the Hubble expansion rate and they should not violate the Big Bang Nucleosynthesis (BBN) bounds \citep{reheating, Berezhiani:2000gw}. This could be achieved if mirror and ordinary worlds are reheated asymmetrically after inflationary epoch. Then the contribution of mirror neutrinos could be suppressed by a factor $x^4$, where
\begin{equation} \label{x}
x = \frac {T'}{T}
\end{equation}
is the temperatures ratio parameter \citep{reheating, Berezhiani:2000gw}. If these two worlds receive different initial temperatures and after that evolve adiabatically, interacting only weakly through gravity, they maintain the initial temperature ratio (\ref{x}) until today \citep{mirror, Berezhiani:2000gw}. BBN bounds require $x<0.5$ \citep{Berezhiani:2000gw}, but stronger limit $x<0.3$ comes from constraints imposed by the cosmological large scale structure formation and cosmic microwave background data \citep{str1, str2}.

In the context of grand unified theories, or electroweak baryogenesis scenario, due to lower temperature, baryon asymmetry in M-World is greater than in our world \citep{Berezhiani:2000gw}. Certain leptogenesis mechanism via common $B-L$ violating interactions between ordinary and mirror particles \citep{leptogenesis, Berezhiani:2008zza}, with $x \lesssim 0.2$, can imply
\begin{equation} \label{omega}
\frac{\Omega'}{\Omega} \approx 5
\end{equation}
($\Omega$ and $\Omega'$ are ordinary and mirror matter densities), which can naturally solve so called coincidence problem between amounts of visible and dark matters and explain all DM in the universe by mirror matter \citep{str3}.

It is known that DM forms spherical halos in galaxies. One may expect that, having the same microphysics as ordinary matter, mirror matter should also form disks, instead of spherical halos. But spherical halos are possible if mirror stars are formed earlier than ordinary stars, and before mirror matter started to collapse into disks \citep{dama}. The recent analyses of strong gravitational lensing caused by DM, revealed that cluster substructures are by more than an order of magnitude efficient lenses than predicted by cold DM simulations \citep{Meneghetti:2020yif}. This also encourages the idea of mirror matter as a cold DM candidate.


\subsection{Mirror stars}

Due to some factors, evolution of mirror stars can be somehow different from ordinary stars. In our world the star formation rate, which is typically adopted from the best-fit-function of experimental data \citep{SFR},
\begin{equation} \label{SFR}
{\rm SFR(z)} = 0.015 \frac{(1+z)^{2.7}}{1+[(1+z)/2.9]^{5.6}} ~ \rm M_{\odot} ~ Mpc^{-3} yr^{-1}~,
\end{equation}
peaks at $z \sim 2$, which corresponds to the lookback time $\sim 9.2~{\rm Gyr}$. However. in M-World with the lower temperature, $T' < T$, all the processes occur earlier at higher redshifts \citep{str3},
\begin{equation}
z' \approx \frac zx~.
\end{equation}
This means that the star formation rate (\ref{SFR}), depending on the temperature ratio (\ref{x}), will peak earlier at $z \sim 10$ \citep{Beradze:2019dzc, Beradze:2019ujd}, corresponding to the lookback time in our world
\begin{equation}
t \sim 13.3 ~{\rm Gyr} .
\end{equation}
This implies that mirror BHs and NSs have more time to pick up mass and to create binaries in the area covered by the LIGO observations.

As M-World is several times colder, at ordinary BBN epoch the universe expansion rate is completely determined by ordinary world itself. So, M-World contribution into the ordinary light element production is negligible \citep{mirror, Berezhiani:2000gw}. In contrary, in the M-World nucleosynthesis epoch, the contribution of ordinary matter scales as $x^{-4}$ and plays a crucial role. It was shown that, for $x \lesssim 0.3$, the mirror helium mass fraction can reach $75-80 \% $ \citep{Berezhiani:2000gw}. Thus, M-World is dominated by mirror helium \citep{mirror} and mirror stars are mostly He-stars \citep{Berezhiani:2005vv}.

Evolution of He-stars should be similar to ordinary stars, when latter have converted most of hydrogen into helium and formed a helium core. During the process of gravitational collapse of protogalaxy, it fragments into hydrogen clouds, which then cools and collapses until the opacity of the system becomes so high that the gas prefers to fragments into protostars. This is a way how first stars (Pop. III stars) in the Universe are formed. The lack of metals for that time, makes cooling process less efficient within clouds. So, their fragmentation could produce only high mass stars.

In the He-dominated world, the cooling process inside primordial clouds should have also a lower efficiency and mirror stars are formed even more massive \citep{Berezhiani:2005vv}. We know that higher is the mass of the ordinary star, the shorter is its life, as it burns out fuel faster. Increasing the initial helium abundance of a star, corresponds to the increase of the mean molecular weight, and correspondingly in both luminosity and effective temperature, that leads to the shorter lifetime. For instance, $10 M_\odot$ star with $70\%$ initial He content has the evolution timescale $\sim 10$ times faster than the star with ordinary He abundance ($24\%$) \citep{Berezhiani:2005vv}.


\subsection{Possible sources of some LIGO events}

First let us consider possible M-World origin of the BHs event GW190521. In principle, BH-BH mergers, which account for the most amount of LIGO events, should not have optical counterparts, so they can be originated from both normal (Pop III) stars and mirror ones. However, BH binaries of mirror origin merely amplifies chance of these BH-BH mergers \citep{Beradze:2019dzc, Beradze:2019ujd}. As the microphysics of mirror stars is similar to that of ordinary stars, they probably also are subject to pair instability and produce the mass gap for intermediate mass BHs. However, stars in M-World are born with higher initial mass, compared to ordinary stars, they evolve faster and higher quantity of massive BHs are formed in short period of time. Adding the fact that the mirror matter density is $\sim 5 $ times the ordinary matter density, collisions of BHs formed by mirror stars are more frequent, increasing merger rate naturally \citep{Beradze:2019dzc, Beradze:2019ujd}. As a consequence, formation of intermediate mass BHs is easier in M-World, that could be a good interpretation for the heavy components of GW190521. Also, BHs formed in the mirror matter environment, can increase in mass by accretion of mirror matter that has higher abundance compared to ordinary matter. In the framework of the M-World scenario, the extreme assumptions made by \citep{Liu:2020gif} in equation (\ref{rate}) may seem more reasonable, and the obtained merger rate for the first generation BHs, that formed the heavy components of GW190521, can look more natural.

Another consequence of the M-World scenario, can be explanation of lower mass gap compact objects of the events GW190425 and GW190814. NS-NS or BH-NS mergers, in case they contain normal NSs, both should be typically accompanied by GRB and optical afterglows. However, neither GW190425 nor GW190814 had such associations \citep{Abbott:2020yvp}. In our paper \citep{Beradze:2019yyp} we suggested that this could indicate to their mirror origin, i.e. as merger of mirror NSs in which case no optical counterpart should be expected. This can be not completely true in the presence of neutron-mirror neutron transitions, which can be rather fast process \citep{Berezhiani:2005hv, Berezhiani:2008bc}. Due to this effect, cores of normal matter can be formed inside the mirror NS \citep{Berezhiani:2020zck}, which can make their merger also optically observable though perhaps more faint.

So, the fact that "heavy" NSs are not detected through electromagnetic spectrum but are observed through gravitational radiation, may be indication that they exist in the mirror world. As discussed above, in order to form a GW190425-like binary NS system, ultra-tight binary with NS and massive He-star is required, that is more easily achieved in mirror world, as M-World is inhabited mostly by He-stars. The formation of GW190814-like systems is also challenging for current theories and their abundance is expected to be extremely low. However, in M-World the abundance of matter exceeds $\sim 5$ times the abundance of ordinary matter and stars in M-World evolve a way faster. This increases the probability of hierarchical mergers may by an order of magnitude, and the formation of GW190814-like systems is more common.


\section{Conclusions}

In our previous papers \citep{Beradze:2019dzc, Beradze:2019ujd, Beradze:2019yyp} we had proposed that most of LIGO events may have emerged from Mirror World. This gives a good explanation for the absence of electromagnetic counterpart radiation and the high merger rates. In this article, we extended that idea on the recently detected unexpected LIGO events. To form binaries similar to GW190521, GW190425 and GW190814, the component masses of which lie in the upper and lower mass gaps, hierarchical mergers of very rare systems are required. We argue that such scenario is order of magnitude more probable in M-World, since it can fully explain DM, i.e. its matter abundance exceed ordinary one about 5 times. Moreover, M-World is dominated by helium; He-stars evolve faster and create compact objects earlier. So, in M-World hierarchical mergers are more probable and second-generation compact objects (remnants of first-generation mergers) are formed with higher rate.


\section*{Acknowledgements}

This work was supported by Shota Rustaveli National Science Foundation of Georgia through the grant DI-18-335.


\section*{Data availability}

There are no new data associated with this article.


\end{document}